\documentclass[twocolumn,english,PRA]{revtex4}
\usepackage[T1]{fontenc}
\usepackage[latin9]{inputenc}
\usepackage{verbatim}
\usepackage{amsmath}
\usepackage{amssymb}
\usepackage{graphicx}
\usepackage{esint}

\makeatletter
\@ifundefined{textcolor}{}
{%
 \definecolor{BLACK}{gray}{0}
 \definecolor{WHITE}{gray}{1}
 \definecolor{RED}{rgb}{1,0,0}
 \definecolor{GREEN}{rgb}{0,1,0}
 \definecolor{BLUE}{rgb}{0,0,1}
 \definecolor{CYAN}{cmyk}{1,0,0,0}
 \definecolor{MAGENTA}{cmyk}{0,1,0,0}
 \definecolor{YELLOW}{cmyk}{0,0,1,0}
}

\usepackage{dsfont}\usepackage{psfrag}\usepackage{epsfig}% Include figure files

\makeatother

\usepackage{babel}
\begin{document}

%\title{Superfluid-Insulator transitions of collective helical modes in the
%zero quantum Hall state of bilayer graphene}
%\title{Superfluid-Insulator transitions of helical quantum Hall Edge modes in bilayer graphene: a realization of quantum spin-ladders}
\title{Helical quantum Hall Edge modes in bilayer graphene: a realization of quantum spin-ladders}

\author{Victoria Mazo}

\affiliation{Department of Physics, Bar-Ilan University, Ramat-Gan 52900, Israel}

\author{Chia-Wei Huang}

\affiliation{Department of Physics, Bar-Ilan University, Ramat-Gan 52900, Israel}

\affiliation{Max Planck Institute for solid state research, Heisenbergstr.1, D-70569,
Stuttgart, Germany}

\author{Efrat Shimshoni}

\affiliation{Department of Physics, Bar-Ilan University, Ramat-Gan 52900, Israel}

\author{Sam T.~Carr}

\affiliation{School of Physical Sciences, University of Kent, Canterbury CT2 7NH,
UK}

\author{H.~A.~Fertig}

\affiliation{Department of Physics, Indiana University, Bloomington, IN 47405,
USA}

\date{\today}
\begin{abstract}
The rich phase diagram of quantum spin-ladder systems has attracted much attention in the theoretical literature.  The progress in experimental realisations of this fascinating physics however has been much slower.  While materials with a ladder-like structure exist, one always has coupling between the ladders to muddy the waters.  In addition, such materials exhibit limited (if any) tunability in terms of the magnetic exchange parameters, and experimental probing of the different phases is a great challenge.  In this work, we show that a realisation of spin-ladder physics can occur
 in an engineered
nanostructure made out of bilayer graphene in the $\nu=0$ quantum Hall state. Specifically, we describe a split-double-gated setup in which a domain wall is explicitly induced in the middle of the sample, and show that an effective spin-ladder forms along this domain wall.  The interaction strengths of the
ladder are tunable by adjusting magnetic and electric fields as well as
the spacing between the gates.  Furthermore, we demonstrate that the effective spin ladder has a helical nature, meaning that the spin-correlations may be probed rather simply with charge transport experiments.  We describe the phase diagram of this system, and show that certain transport measurements are very sensitive to the phase.
\end{abstract}


\maketitle
%73.21.-b: Electron states and collective excitations in multilayers, quantum wells, mesoscopic, and nanoscale systems; 73.22.Gk: Broken symmetry phases;
%73.43.Lp: Collective excitations (in context of QHE); 72.80.Vp: Electronic transport in graphene; 75.10.Pq: Spin chain models;
%75.10.Jm: Quantized spin models, including quantum spin frustration

\section{\label{sec:-Introduction} Introduction}

It has been known for many years that the properties of one-dimensional Heisenberg spin-chains depend in a crucial way on whether the spin of the system is integer or half-integer \cite{Haldane}.  While half-integer spin chains have gapless excitations and power-law correlations, those with integer spin demonstrate a spin-gap and spin-liquid behavior.  Such arguments were later extended to $S=1/2$ spin-ladders, consisting of coupled $n$ legs \cite{Dagotto1,Rice1,Rice2}, in this case the difference being between an even and odd number of legs.

The spin-$1/2$ two-leg spin ladder has since become one of the most studied problems in low-dimensional quantum magnetism \cite{Giamarchi,Gogolin}, as it is in some sense intermediate between the spin-$1/2$ and spin-$1$ systems, thus leading to a better understanding of the Haldane result \cite{Haldane}.   The Hamiltonian
\begin{multline}
\label{eq: conf ham}
H=\sum_{j,n=1,2}\{J_{\parallel}^{xy}(S^+_{j,n}S^-_{j+1,n}+h.c.)+J_{\parallel}^zS^z_{j,n}S^z_{j+1,n}\} \\
+
\sum_{j}\{J_{\perp}^{xy}(S^+_{j,1}S^-_{j,2}+h.c.)+J_{\perp}^zS^z_{j,1}S^z_{j,2}\}
\end{multline}
%{[}as expressed in Eq. (\ref{spinladder}){]}
describes two spin-$1/2$ chains on the legs $n=1,2$ with in-chain couplings $J_\|$, coupled via rung coupling constants $J_\perp$.  For more generality, we have also allowed each of the exchange couplings to have some XXZ anisotropy.  However initially we may consider Heisenberg exchange $J^{xy}=J^z$.  We then can see if $J_\perp=0$, we have two decoupled spin-$1/2$ chains; while if $J_\perp$ is large and ferromagnetic in sign, then the two-spins on each rung combine to form a triplet state, giving an effective spin-$1$ chain, which should thus exhibit the Haldane gap.  On the other hand if $J_\perp$ is large and antiferromagnetic, then one finds a gapped system of rung-singlets.

The question remains of what happens at small $J_\perp$.  This was solved by Shelton et al. \cite{Shelton} through bosonization and refermionization, where they showed that a spin-gap opens up immediately at $J_\perp \ne 0$ for either sign. In this limit, the gap is associated with confinement of spinons and most interestingly the structure of the theory is independent of the sign of $J_\perp$.  However it was later shown \cite{Lecheminant,Orignac} that the response to this system to boundaries (or impurities) is rather different in the two phases: for ferromagnetic rung couplings one finds spin-$1/2$ states localised on the boundaries, while these edge-states are absent if the rung couplings are antiferromagnetic.  In modern day terminology, one can therefore say that $J_\perp=0$ is a transition between a topological phase and a non-topological one.

While this phase transition at $J_\perp=0$ is arguably the most discussed in the physics of spin-ladders, a number of additional transitions are possible when the XXZ spin anisotropy is taken into account \cite{Giamarchi}; we will discuss these transitions in detail later in this work.  Other investigations have concentrated on the case when the legs are ferromagnetic \cite{Japaridze}, leading to further interesting transitions.

To date, there are a few ladder compounds, for example, vanadyl pyrophosphate
(VO)$_{2}$P$_{2}$O$_{7}$ and the cuprates series Sr$_{n-1}$Cu$_{n+1}$O$_{2n}$,
which have been identified in experiments and the spin gaps have been measured
by neutron scattering or NMR \cite{Dagotto}. More recently, the organic compound ${\rm Br_4(C_5H_{12}N)_2}$ (BPCB)
has been studied \cite{BPCB}, where the relatively low exchange coupling enables closing of the gap with a magnetic field.  However,
despite those breakthrough experiments, investigation of the phase-diagram and nature of excitations of spin-ladder systems remains challenging for three main reasons.

Firstly, all examples in solid-state physics have inter-ladder couplings.  These become important at low temperatures, and make it harder to separate contributions from the one-dimensional ladder physics and those from the three-dimensional nature of the true crystal.  While in certain situations, the one-dimensional physics may still be dominant, even within the three-dimensional ordered phase \cite{SamOld}, it is clearly desirable to isolate a single spin ladder, both in terms of nanotechnology as well as from the fundamental physics point of view.

Second, the interaction strengths
$J_{\bot}$ and $J_{\Vert}$ in the ladder compounds are not tunable, and are fixed by the chemistry of the compound in question.  Therefore, the quantum phase transition predicted by the theory can not be simply accessed in real materials.  We should note however that recent advances in chemical doping have gone some way towards solving this problem \cite{Zheludev}.

Thirdly, to probe the spin-correlations requires techniques such as neutron scattering, which are bulk measurements and therefore not appropriate to a single-spin ladder, should one be isolated.  This is in strong contrast to Fermionic ladder systems, which have an equally rich phase diagram, but which may be probed by charge transport \cite{Sam_long}.  An (undoped) spin-ladder is a Mott insulators and the charge sector is completely frozen; all low energy excitations are in the spin sector, and therefore do not contribute to electric transport.  Other transport measurements such as thermal conductivity \cite{thermal_exp} are possible and may yield signatures of the different spin phases, however the spin contribution is difficult to disentangle from that due to crystal phonons \cite{spin_phonon}.

Many of the challenges to probing these systems and their phase diagrams
can be overcome by considering systems whose low-energy dynamics are isomorphic
to those of spin ladders.  Of such systems, graphene has emerged as a
particularly useful paradigm.  In the quantum Hall regime, single layer
graphene systems can have edge states with definite spin that (when the
system is undoped) cross the Fermi level and one
another \cite{BreyFertig,Abanin_2006}, forming an effective one-dimensional system
in which charge and spin degrees of freedom are locked together, and
in which interaction effects can lead to highly non-trivial phases and
associated transport properties which can usefully be modeled as a
spin-chain system \cite{fertig_2006,SFP}.  By putting together copies
of such systems, one may create what is effectively a spin-ladder
system at low energies.

Such systems naturally occur in bilayer graphene, either at its edge \cite{Mazo,Mazo2}
or within its bulk \cite{Huang, Mazo-1}, using a split gate geometry first introduced
in  Ref. \onlinecite{Martin2008} as a zero-field system.
Recently, we have proposed \cite{Mazo-1} an experimental realization of this which allows
for engineering a spin ladder with tunable interactions, and enables it to be probed in {\it charge transport} measurements  by exploiting the spin-charge coupling that
is inherent to the system. Distinct phases in the spin sector are therefore
manifested by distinct electric conduction properties.

In these proceedings, we review this proposal, as well as present some new results on the charge transport behavior.  The paper is organised as follows.  In Section \ref{sec:system}, we introduce our proposed setup and show that appropriate gating of bilayer graphene can induce an effective spin-ladder system. We then give the Hamiltonian of this spin-ladder system paying particular attention to the relationship between the model parameters and the original engineering setup.  In Section \ref{sec:Phase-diagram} we solve this model, deducing the phase diagram of the spin-ladder system.  In Section \ref{sec:Transport-coefficients}, we then demonstrate that the different phases of this system lead to dramatically different conductance properties -- in particular there is a transition between a density-wave insulator and a superconducting phase.  Finally, the manuscript ends with some concluding remarks.  Throughout the paper we use units in which $\hbar=1$.

%%%%%%%%%%%%%%%%%

\section{Model system and Hamiltonian}
\label{sec:system}

\begin{figure}
%\vspace{-10pt}
%\scalebox{0.09}{{\includegraphics{dgated_BLG}}}
\begin{centering}
\includegraphics[width=1\linewidth]{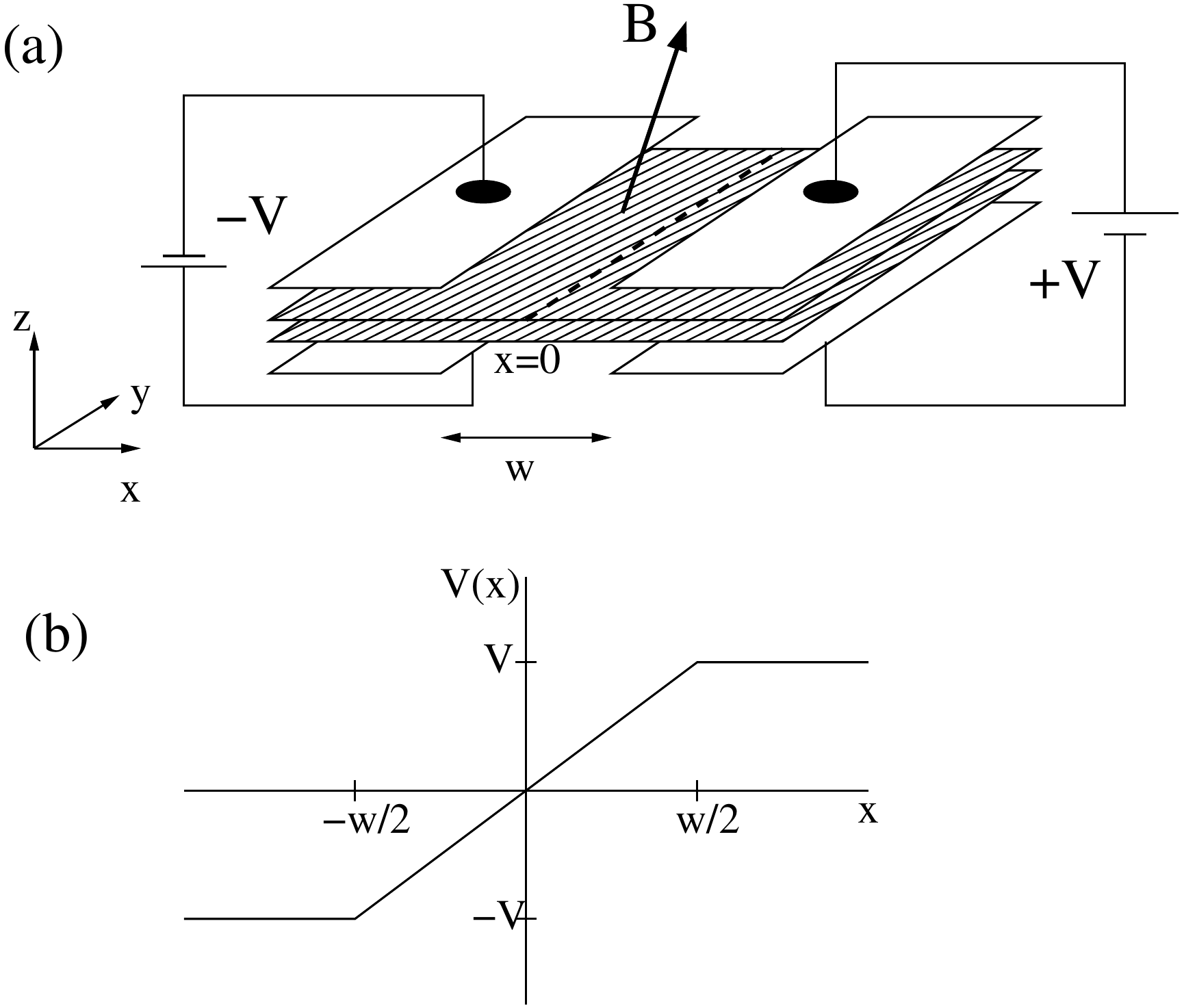}
 \caption{\label{fig:dgated_BLG} (a) Our proposed setup consisting of BLG in a magnetic field and subject to a split-double-gate creating an inhomogenous electric field.   The magnetic field ${\bf B}$ is tilted away from the $z$ direction, allowing the magnetic length and Zeeman splitting to be controlled independently.  The line $x=0$ is the location of the effective one-dimensional modes that form the ladder model. (b) A schematic of the inhomogenous potential across the sample.  The gradient of the potential near $x=0$ can be controlled by varying both $V$ and the distance between the split gates $w$. }
\par\end{centering}
%\vspace{-10pt}
\end{figure}

Our proposed setup consists of a sheet of
bilayer graphene (BLG) subject to a magnetic field and a split-double
gate, as depicted in Fig. \ref{fig:dgated_BLG}(a) \cite{Martin2008}.  The gates are arranged in such a way that there is a perpendicular electric field applied to the sample, that changes sign in the middle.  This implies that there is a line in the middle of the sample at $x=0$ where the perpendicular electric field vanishes -- see Fig.~\ref{fig:dgated_BLG}(b).  We will now demonstrate that if the BLG is undoped, the bulk of the sample exhibits an excitation gap, but there exists one-dimensional low-energy modes within this gap localised near the region $x=0$ where the electric field changes sign.  Furthermore, we will show that these low-energy modes can be described by an effective model of two weakly-coupled spin ladders.  The parameters of the spin-ladders can be tuned by varying the magnetic and electric fields, and furthermore the spin-ladders have a helical structure, meaning that the spin excitations may be probed by charge transport experiments.

This mapping was first described in Ref.~\onlinecite{Mazo-1}.  In these proceedings, we give a much more pedagogical account of the mapping to the spin-ladder; we believe that this is useful to aid understanding of the original derivation of this mapping \cite{Mazo-1}.

\subsection{From bilayer graphene to spin-ladder: noninteracting picture}

\begin{figure}
%\vspace{-10pt}
%\scalebox{0.09}{{\includegraphics{dgated_BLG}}}
\begin{centering}
\includegraphics[width=0.8\linewidth]{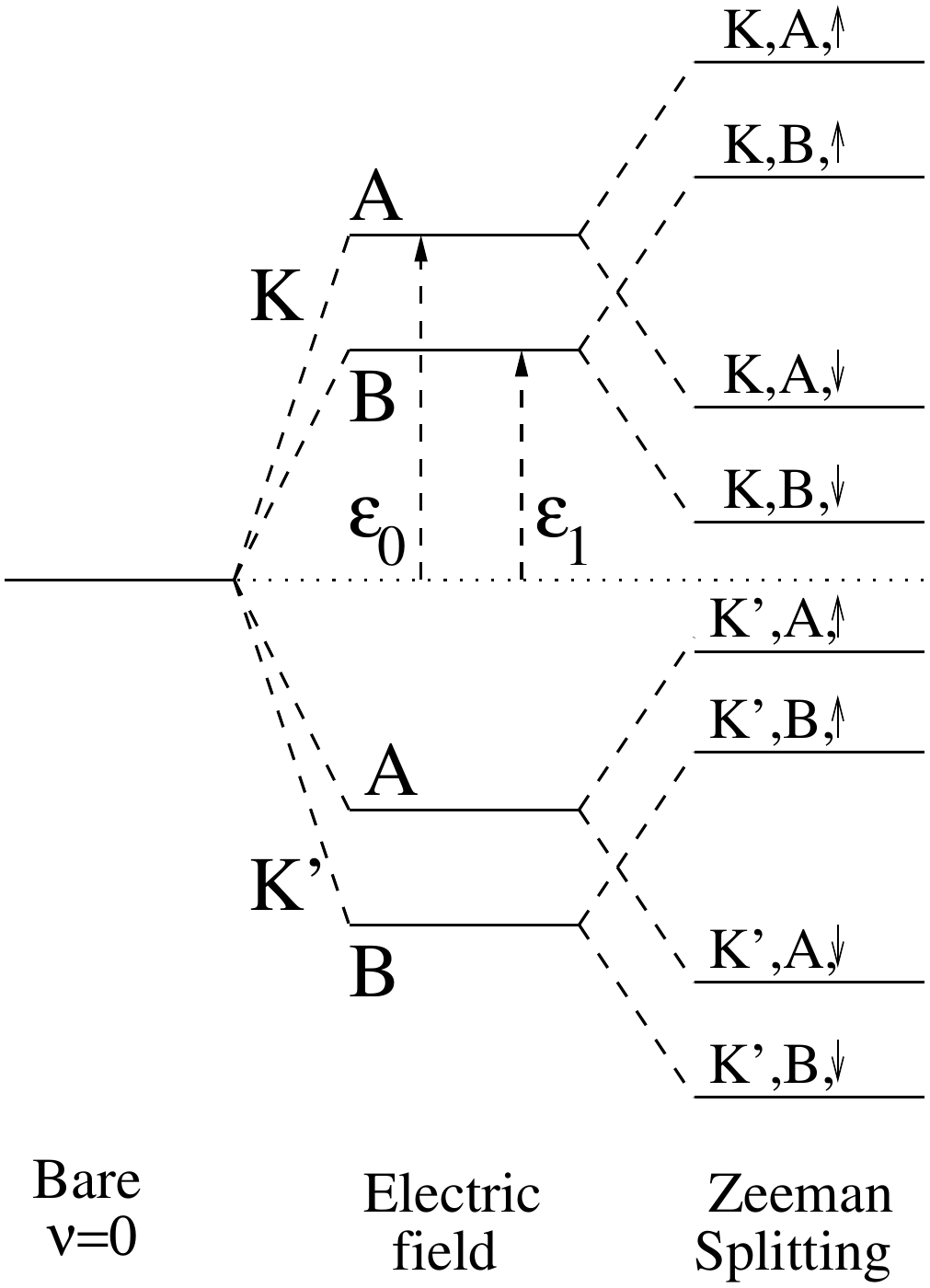}
 \caption{\label{fig:LevelSplitting} The splitting of the $\nu=0$ Landau level in BLG.  In the bare model of the Landau levels, this level at energy $\epsilon=0$ is $8$ fold degenerate, coming from the two layers, the two $K$ points, and the two spin projections.  In the addition of an electric field perpendicular to the BLG, this is split into four levels at energies $\epsilon=\pm \epsilon_1$ and $\epsilon-\pm \epsilon_2$ where both $\epsilon_{1,2} \propto V$.  For the sign of $V$ plotted, the upper two levels come from the $K$ point while the lower two are from the $K'$ point.  The further splitting involves states mostly localised on the upper (A) or lower (B) layers of the BLG.  Finally, adding the Zeeman splitting term (which is independent of the potential $V$) further splits each of these levels into two.  It is clear that the ultimate arrangement of the levels depends on both the sign and magnitude of $V$.}
\par\end{centering}
%\vspace{-10pt}
\end{figure}

\begin{figure}
%\vspace{-10pt}
%\scalebox{0.09}{{\includegraphics{dgated_BLG}}}
\begin{centering}
\includegraphics[width=1\linewidth]{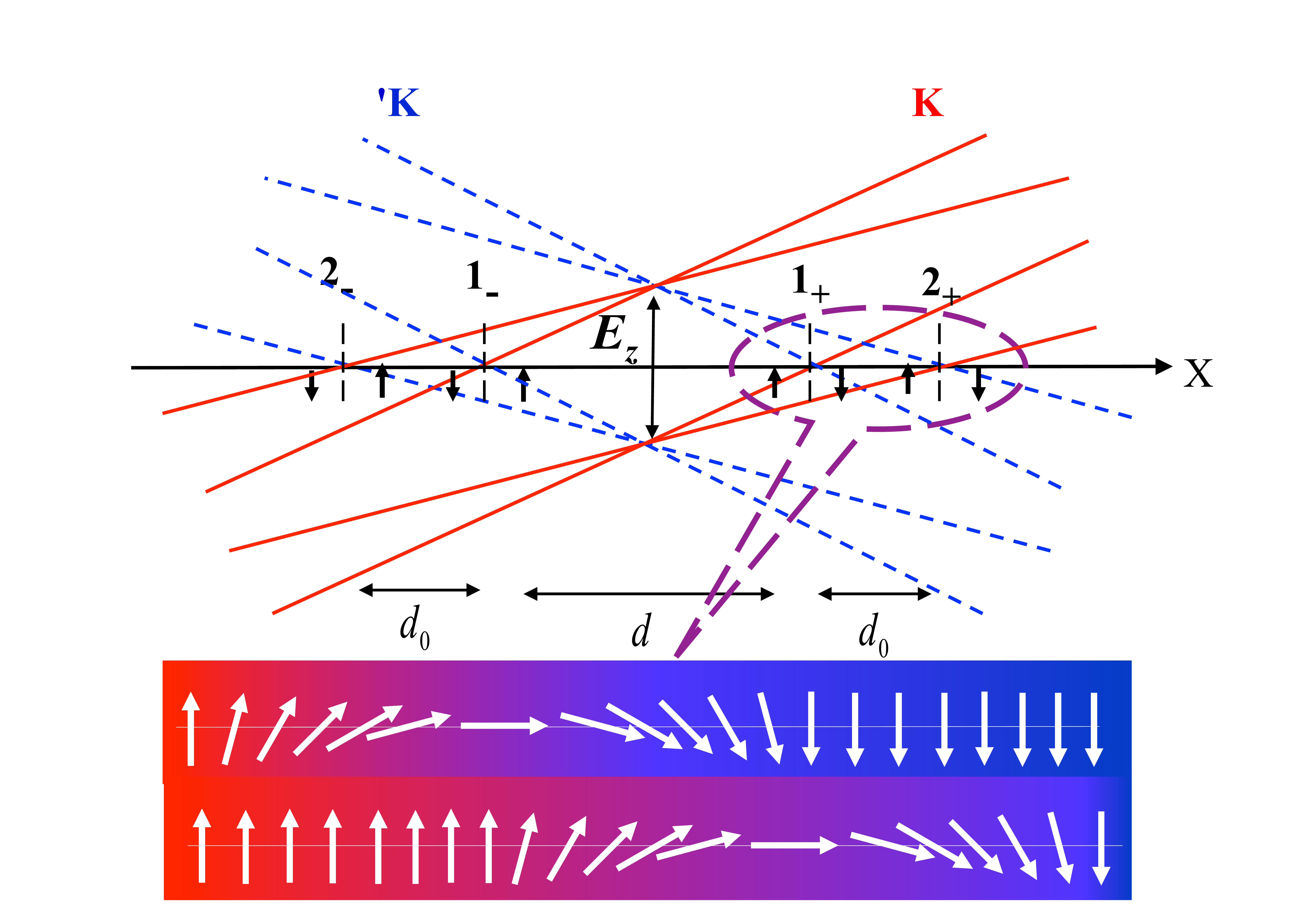}
 \caption{\label{fig:Crossings}(color online) The non-interacting
energy levels crossing at the undoped Fermi energy, $\epsilon=0$ as a function of the position of the guiding center $X$. Arrows
denote spin projection ($S^{z}$ along ${\bf B}$), full red lines correspond
to valley $\mathbf{K}$ and dashed blue lines to valley $\mathbf{K'}$.  The different slopes correspond to the different levels $\epsilon_{0,1}$.
The blow-up presents a typical $S^{z}$-configuration on the two layers as one scans through a pair of domain walls.}
\par\end{centering}
%\vspace{-10pt}
\end{figure}

We begin by discussing the non-interacting system.  The Landau quantisation of BLG subject to a perpendicular magnetic field $B_z$ has been widely discussed \cite{McCann2006,Mazo,Huang}, where one finds that the lowest Landau levels are given by an energy
\begin{equation}
\epsilon_n = \pm \frac{\omega_c^2}{\gamma_1} \sqrt{n(n-1)}, \;\;\; n=0,1,\ldots
\label{eq:LL}
\end{equation}
This equation requires a few remarks.  The cyclotron frequency
 \begin{equation}
 \omega_{c}=\sqrt{2}\hbar v_{F}/\ell
 \label{eq:wc}
 \end{equation}
(here $\ell=\sqrt{c\hbar/eB_z}$ is the magnetic length) is defined in terms of the high energy spectrum of bilayer graphene which is linear \cite{McCann2006} in the absence of the magnetic field.  In contrast, the low energy spectrum becomes quadratic in the presence of the interlayer hopping $\gamma_1$, leading to an effective low-energy cyclotron frequency $\tilde{\omega}_c=\omega_c^2/\gamma_1$.  This inessential detail aside, the most curious thing about Eq.~\eqref{eq:LL} is the fact that the lowest two Landau levels $n=0$ and $n=1$ are degenerate.  Combining this with usual valley and spin degeneracies of graphene means that the lowest Landau level $\nu=0$ is actually eight-fold degenerate.

This degeneracy is lifted when a potential difference $2V$ is applied perpendicular to the BLG sample, leading to an imbalance in potential between the upper and lower layers.  The previously degenerate levels $\epsilon_0=\epsilon_1=0$ then become split with energies \cite{McCann2006,Mazo,Huang}
\begin{align}
\epsilon_0 &= \pm \frac{eV}{2} \nonumber \\
\epsilon_1 & = \pm \frac{\gamma_1^2-\omega_c^2}{\gamma_1^2+\omega_c^2} \frac{eV}{2}.
\end{align}
This splitting is illustrated in Fig.~\ref{fig:LevelSplitting}.  When the wave functions are calculated, it is found that the levels with the $+$ signs are predominantly states from the $K$ valley while those with the $-$ sign come from the $K'$ valley.  The further splitting between $\epsilon_0$ and $\epsilon_1$ can be further associated with states located on the upper or lower layers of the BLG in the manner shown in Fig.~\ref{fig:LevelSplitting}.  It is worth pointing out here that for realistic magnetic fields, $\omega_c/\gamma_1 \ll 1$, which means that the splitting between the $\epsilon_0$ and $\epsilon_1$ levels, $(\epsilon_1-\epsilon_0) \sim (\omega_c/\gamma_1)^2 eV$ is much smaller than the splitting between the plus and minus states $(\epsilon_{0+}-\epsilon_{0-}) = eV$.  However, each of these splittings is proportional to $V$, so all go to zero as $V\rightarrow 0$.

Finally, the spin degeneracy of each of these four levels is split by the Zeeman energy $E_z \propto |\mathbf{B}|$.  Notice that this term contains the full magnetic field $\mathbf{B}$ and not just the $z$-component as appears in the cyclotron frequency, Eq.~\eqref{eq:wc}.  This means that these two energy scales can be controlled separately by changing the tilt angle of the magnetic field.  It also means that the spin projection, which we call $S_z$ for convenience, is actually the projection along the direction of $\mathbf{B}$ and not along the true $z$-axis. This simplifies notation and makes no difference to the end results. The combined effect of the layer anisotropy and the Zeeman splitting removes all of the  degeneracies of the original $\nu=0$ level -- see Fig.~\ref{fig:LevelSplitting}.

We see therefore that in a uniform perpendicular electric field
the Hamiltonian for non-interacting electrons in
undoped BLG (chemical potential $\mu=0$) has an energy gap to all excitations.  In the inhomogeneously gated system of Fig.~\ref{fig:dgated_BLG} however, one expects to find
energy crossings in the vicinity of the point $x=0$ where
the local electric potential difference $V_X$ is small, and therefore one expects to see one-dimensional gapless modes delocalised in the $y$-direction, but localised in the $x$-direction around these crossings.  The first such crossing occurs when the $\epsilon_{0+}$ state with $s_z=-1$ crosses the $\epsilon_{0-}$ state with $s_z=+1$, i.e.,
\begin{equation}
eV/2-E_z=-eV/2+E_z=0 \;\;\; \implies eV=2E_z.
\end{equation}
Assuming that the potential around $x=0$ is given by $V_X=2\frac{V}{w}X$ (see Fig.~\ref{fig:dgated_BLG}(b)), this implies a crossing at
\begin{equation}
X_1=\frac{E_z}{2eV/w}.
\end{equation}
We mention that in the context of the inhomogenous system, the coordinate $X$ is best thought of as the coordinate of the guiding centre of the individual states within the Landau level.

One similarly finds a crossing at $-X_1$, as well as the crossings where the $\epsilon_1$ levels cross: $X=\pm X_2$ where
  \begin{equation}
  X_2=
\frac{E_z}{2eV/w} \frac{\gamma_1^2-\omega_c^2}{\gamma_1^2+\omega_c^2}.
\end{equation}
This is illustrated in Fig.~\ref{fig:Crossings}.  Here, it can be seen that the crossings, which correspond to low-energy effective one-dimensional modes, naturally arrange themselves in two groups of two.  The splitting between the two groups is
\begin{equation}
d\ = 2X_1 = w\frac{E_{z}}{eV},\label{eq:d}
\end{equation}
while the splitting between the two legs in each group is given by
\begin{equation}
d_{0} = X_2-X_1 = d\frac{\omega_{c}^{2}}{\gamma_{1}^{2}-\omega_{c}^{2}}.
\end{equation}
We therefore see that in the realistic limit when $\omega_c \ll \gamma_1$, we have $d_{0} \ll d$; this implies that when interactions between these modes are taken into account, one may consider them as two weakly-coupled two-leg ladders.

The final thing to discuss concerns the nature of the modes at each of these crossing points.  The $\nu=0$ state of either monolayer graphene of BLG is unusual in that it contains both particle and hole states.  This means that at a boundary \cite{BreyFertig,Abanin_2006,Mazo} or an artificial edge \cite{Martin2008,Paramekanti,Huang}, edge states of both chiralities are formed.  It turns out \cite{Mazo,Huang} that when the degeneracy is lifted by adding the electric field, the particle states reside in one valley, while the hole states reside in the other.  This means that the valley states are chiral, and therefore the induced one-dimensional boundary modes have a chirality index $\mu=\pm 1$ equal to their valley index.  As the modes are also further split into their different spin projections, every mode can be assigned a helicity index $h=\mu s_z$.

\begin{figure}
\includegraphics[width=1\columnwidth]{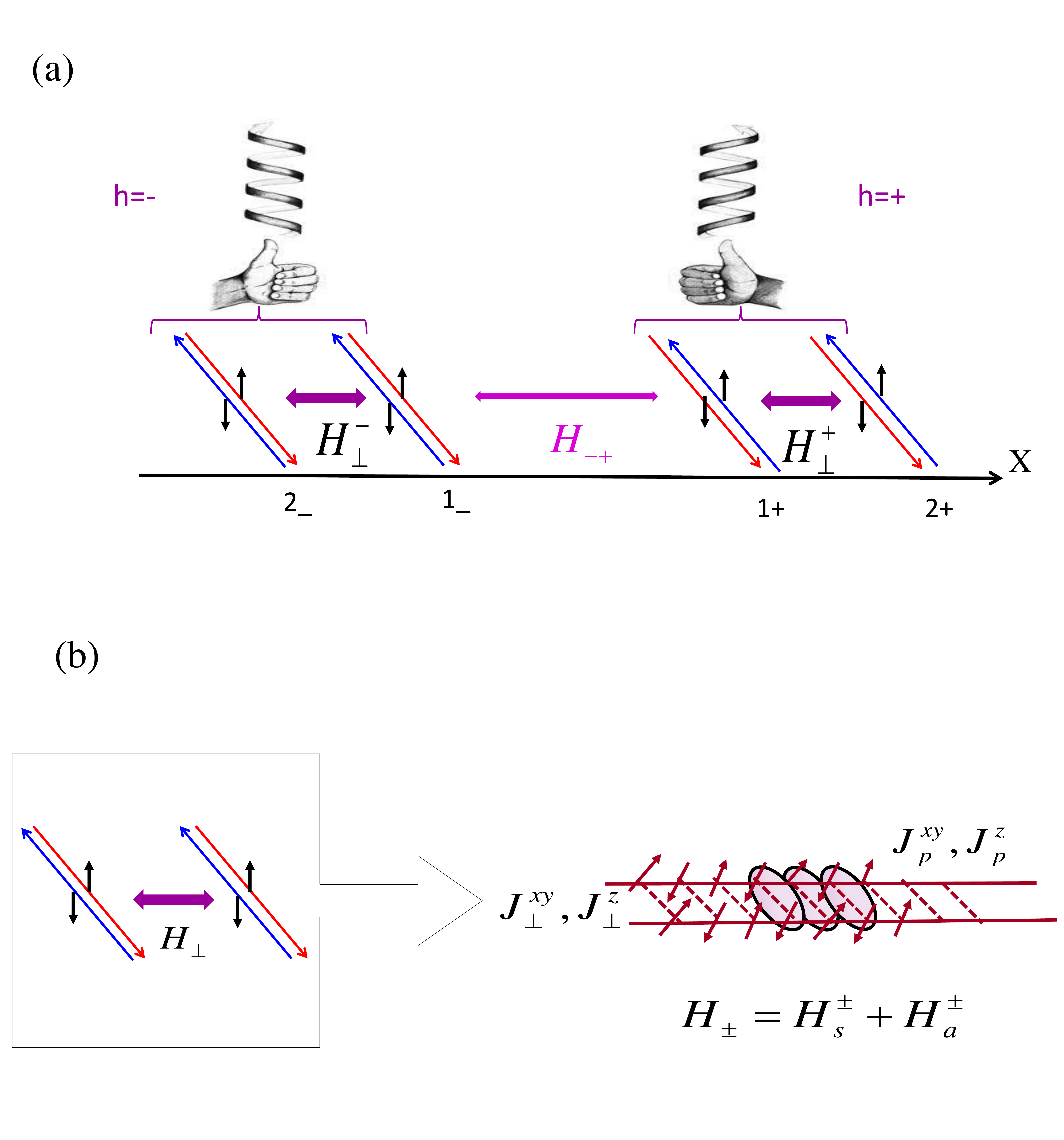}\caption{\label{fig:hamiltonian}Pictorial representation of two coupled spin$1/2$
ladders. (a) The blow-ups in Fig. \ref{fig:dgated_BLG}b is represented
as counter propagating spin currents, and they are mapped to spin
chains. Depending on the choices of the fields and the physical spacing
between the split gates, one could tune the system in such a way that
the chains with the same helicity coupled stronger than that with
different helicity. Consequently, the chains with the same helicity
are coupled to form a spin ladder which is weakly coupled to another.
(b) Each spin 1/2 chain is represented as the standard XXZ Hamiltonian.
When one is coupled to the other by the interaction $J_{\perp}$,
a two leg ladder forms. }
\end{figure}

The curious thing, that can be determined by direct calculation or by simply studying Fig.~\ref{fig:Crossings} is that the four modes with guiding centers located at $X>0$ all have one helicity (call it $h=+1$), while those at $X<0$ have the opposite helicity ($h=-1$).  This is depicted schematically in Fig.~\ref{fig:hamiltonian}(a).  Thus one has two weakly coupled two-leg ladder models, with the two copies with opposite helicities only weakly coupled to each other.  Each leg has two modes of opposite chirality as one finds in usual single-channel one-dimensional systems \cite{Giamarchi}, however each with the same helicity.  This means that one can write the modes either in terms of charge currents or in terms of spins, the two being linked.  With interactions, the spin representation turns out to be the most convenient, as we now discuss.

\subsection{Interactions and effective low energy Hamiltonian}
\label{sec:Model-Hamiltonian}

The scenario we have described above is based on a model of non-interacting electrons.  However as the Landau levels are flat bands and thus have quenched kinetic energy, Coulomb interactions can have a dramatic effect on the nature of the ground state and excitations.  One example of this is the spontaneously polarised ferromagnetic state, which has been extensively discussed in the context of ordinary quantum Hall systems in two-dimensional electron gases \cite{QHFM}.  In this case, interactions lift the spin-degeneracy of the Landau-levels by forming a symmetry-broken state.  Related phenomena have been predicted and observed in the $\nu=0$ state of graphene (both mono-layer and bilayer) \cite{CastroNetoRMP,Abanin2007,Checkelsky,Du2009,Feldman2009,Jung2009,Nandkishore,Kharitonov_bulk,QHFMGexp}; in this case the additional discrete degrees of freedom (valley, and layer index in BLG) lead to a very rich phase diagram of potential broken symmetry states.  It has been argued that one way to experimentally tune between different symmetry-broken phases is by tuning the Zeeman energy via a strong magnetic field in the plane of the graphene  \cite{Young2013}.  Such ideas can be extended to the isospin in BLG (associated with the valley and layer indices) via a perpendicular electric field which acts as an effective Zeeman term \cite{McCann2006,Yacoby2011}, and therefore may be used to experimentally induce different ground states.

One of the defining properties of the symmetry-broken quantum-Hall states is the nature of their collective excitations\cite{QHFM,BreyFertig,Abanin_2006,Mazo}.  In particular, the elementary charge excitations in the standard quantum-Hall ferromagnet are Skyrmions \cite{QHFM}, whereas in graphene, more complex excitations are possible \cite{2eSkyrm,Kharitonov_edge}.  While the Skyrmion in this context is defined as a particular spin-texture, the nature of the spin-charge coupling in quantum-Hall ferromagnets means that the \textit{topological charge} associated with this spin-texture can be equated to the \textit{electrical charge} quantum
number of the Skyrmion.  More generally, this spin-charge coupling phenomenon \cite{QHFM}
opens the possibility of probing spin physics through electrical measurements.

This physics may also be applied at boundaries, whether it is near the physical edge of a graphene ribbon \cite{BreyFertig,Abanin_2006,Mazo}, or an edge induced by a split gate
setup as in Fig.~\ref{fig:dgated_BLG} \cite{Martin2008,Paramekanti,Huang}.  The end result is that level-crossings of the non-interacting picture become a coherent domain wall (DW) (see Fig.~\ref{fig:Crossings}), in which gapless collective excitations may move along this wall.  Residual interactions between these furthermore lead to interactions between the four legs that were originally decoupled in the non-interacting picture.

Putting this on a firm theoretical footing,
the quantum dynamics of %spin fluctuations in
each DW-mode (along the $y$-direction) can be described in terms
of an effectively 1D spin-$1/2$ field (${\bf S}_{n_{h},y}$), simultaneously encoding
both the spin and charge degrees of freedom \cite{BreyFertig,SFP,Mazo2}.
The index $n_h$ takes values $n=1,2$ and helicity $h=\pm$, and refers to the
same four channels that were found in the non-interacting case (see Fig.~\ref{fig:Crossings}).
The spin operator acts on the real spin, but furthermore the $S^{x}$ and $S^{y}$ operators are
also associated with electric charge, while the
the $S^{z}$ operator coincides with the electric \textit{current},
 $S_{n_{h}}^{z}\sim hj_{e}$ (note the role of helicity $h$ in this relation).

The system is therefore modelled by the effective
Hamiltonian (see Fig.~\ref{fig:hamiltonian}(b))
\begin{equation}
H=\sum_{h=\pm}H_{h}+H_{+-}, \label{2ladders}
\end{equation}
where $H_{(h)}$ describe anisotropic spin-$1/2$ two-leg ladders:
\begin{eqnarray}
H_{h} & = & \sum_{n=1,2}H_{n_{h}}+H_{\perp}^{(h)}\;,\label{spinladder}  \nonumber \\
H_{n_{h}} & = & \sum\limits _{y}\left[\frac{J_{n}^{xy}}{2}\big(S_{n_{h},y}^{+}S_{n_{h},y+1}^{-}+h.c.\big)+J_{n}^{z}S_{n_{h},y}^{z}S_{n_{h},y+1}^{z}\right]\label{eq: spinchain}  \nonumber \\
H_{\perp}^{(h)} & = & \sum\limits _{y}\left[\frac{J_{\perp}^{xy}}{2}\big(S_{1_{h},y}^{+}S_{2_{h},y}^{-}+h.c.\big)+J_{\perp}^{z}S_{1_{h},y}^{z}S_{2_{h},y}^{z}\right],\label{eq:transverse_SC}
\end{eqnarray}
and $H_{+-}$ is the weak coupling between the two ladders.  For the purposes of this section, we will assume this coupling to be negligible and study the phase diagram of each individual ladder; this term will be added back in perturbatively later when we discuss transport.  We therefore see that the effective low-energy theory of our double-gated BLG is equivalent to two copies of the original spin-ladder Hamiltonian \eqref{eq: conf ham}.  As we will see now, the parameters of this effective model are very tuneable by simply varying gate voltages and the magnetic field.

The in-chain constant $J_n^{xy}$ comes purely from Coulomb interactions,
and is a combination of a direct interaction due to the spin-charge coupling,
and the spin-stiffness of the underlying two-dimensional electron gas.
The constant $J_n^z$ by contrast comes from
a combination of the electric field gradient within the split gate
and the interaction-induced spin stiffness.
On the other hand, the inter-chain exchange constant $J_\perp^{xy}$ comes
largely from the interaction between the modes localised on each chain
due mostly to the charge induced on them when an in-plane spin gradient
is present \cite{fertig_2006},
while $J_\perp^z$ comes from exchange interactions.  The most important thing here is that the hybridisation term $J_\perp^{xy}(S_{1_{h},y}^{+}S_{2_{h},y}^{-}+h.c.)$ is very different from how the term would look in the original fermion picture -- the correct form for this term is perhaps the most crucial result of the reconstruction of the one-dimensional modes through the bulk interactions.

 The dependence of the model parameters $J_{n}^{\alpha},J_{\perp}^{\alpha}$ on
the original system parameters and external fields is non-universal and extremely complicated.  All of the parameters depend on the details of the overlaps between the different skyrmion wavefunctions, and as such even their signs can not be deduced from general arguments alone.  However despite this, several useful remarks can be made relating to the experimental tuning of the system.  First, the anisotropy factors
\begin{equation}
 \Delta_{n(\perp)}\equiv\frac{J_{n(\perp)}^{z}}{J_{n(\perp)}^{xy}}
 \label{eq:anisotropy}
\end{equation}
qualitatively reflect the ratio of  kinetic energy ($\propto eV$) to exchange interaction ($\sim e^{2}/\ell$),
 and thereby may be varied by changing gate voltages or the magnetic length (i.e. $B_z$).  Secondly, the  magnitude of the inter-chain couplings $J_\perp$ is strongly dependent on the distance between the chains $d$ (see Eq.~\ref{eq:d}).  The ratio between in-chain and inter-chain couplings can therefore be controlled by varying $d$, which from \eqref{eq:d} amounts to changing either the Zeeman energy (i.e. changing $|\mathbf{B}|$), or changing the spatial variation of the gate voltage $V/w$.

We next employ standard Bosonization to express the spin operators
in terms of Bosonic fields $\phi_{n_{h}}(y)$ and their dual $\theta_{n_{h}}(y)$
\cite{Giamarchi}
\begin{align}
 & S_{n_{h},y}^{+}= \sqrt{\frac{\alpha}{2\pi}} \frac{(-1)^{y}}{\alpha} e^{-i\theta_{n_{h}}(y)} \label{eq:JWtransformation}\\
 & S_{n_{h},y}^{z} = \frac{1}{\pi}  [-\partial_{y}\phi_{n_{h}}(y)+\frac{(-1)^{y}}{\alpha}\cos(2\phi_{n_{h}}(y))],\nonumber
\end{align}
where $\alpha \sim \ell$ is the short-distance cutoff, and should also be associated with the lattice spacing of the original spin-ladder model.  The equivalence is exact for non-interacting systems. When interactions are taken into account these two length scales may differ but this is an inessential complication and we don't keep track of this difference here. Keeping only the
most relevant terms, the Hamiltonian in Eq.~\ref{spinladder}
can then be expressed as follows.  First, the in-chain part
\begin{eqnarray}
H_{n_{h}} & = & \frac{u}{2\pi}\int\mathtt{d}y\left[K\left(\nabla\theta_{n_{h}}\right)^{2}+\frac{1}{K}\left(\nabla\phi_{h}\right)^{2}\right]\nonumber \\
 &  & + J_n^{z}  \int  \frac{\mathtt{d}y}{2\pi \alpha^2} \cos4\phi_{n_{h}}.
\label{hnh}
\end{eqnarray}
In this expression, the renormalised velocity $u=J_n^{xy}\alpha/K$, while the Luttinger parameter $K$ may be deduced from Bethe ansatz results to be \cite{Giamarchi}
\begin{equation}
K=\frac{\pi}{2\arccos(-J^z_n/J^{xy}_n)},
\label{Kdef_Delta}
\end{equation}
where we have made the reasonable assumption that the coupling parameters for each of the two legs are equivalent. In fact, a weak anisotropy between the legs is unimportant (see Ref.~\cite{Mazo-1}).
For the XY model, when $J^z=0$, we find that $K=1$; for \textit{antiferromagnetic} interactions $J^z>0$, then $K<1$; while for \textit{ferromagnetic} interaction, then $K>1$.  The cosine term is only relevant for $K<1/2$ (corresponding to an easy-axis antiferromagnet, and beyond the scope of Eq.~\ref{Kdef_Delta}).  We will not be interested in this case here, and will therefore drop the cosine term in Eq. \ref{hnh} from here on.

The interchain interaction may also be bosonized, yielding
\begin{multline}
H_{\perp}^{(h)} = \int\frac{\mathtt{d}y}{2\pi \alpha} \left[J_{\perp}^{xy}\cos\left(\theta_{1_{h}}-\theta_{2_{h}}\right)+J_{\perp}^{z}\cos2\left(\phi_{1_{h}}-\phi_{2_{h}}\right)\right.  \\
  \left.+J_{\perp}^{z}\cos2\left(\phi_{1_{h}}-\phi_{2_{h}}\right)\right]+ J_{\perp}^{z}\alpha \int\mathtt{d}y\,\partial_{y}\phi_{1_{h}}\partial_{y}\phi_{2_{h}}.
\end{multline}

In order to treat this, we define the symmetric and antisymmetric modes in each ladder \cite{Giamarchi} (for later convenience, we adopt a non-canonical form of these rotations involving an extra scaling of the fields)
\begin{align}
\label{eq:rotation}
 & \phi_{s_{h}}=\frac{1}{2}(\phi_{1_{h}}+\phi_{2_{h}})\qquad\quad\,\,\theta_{s_{h}}=(\theta_{1_{h}}+\theta_{2_{h}})\\
 & \phi_{a_{h}}=(\phi_{1_{h}}-\phi_{2_{h}})\qquad\qquad\theta_{a_{h}}=\frac{1}{2}(\theta_{1_{h}}-\theta_{2_{h}}).\nonumber
\end{align}
Executing this rotation, the Hamiltonian decouples into the sum of symmetric and antisymmetric modes
\begin{equation}
H_{h}=H_{s}^{(h)}+H_{a}^{(h)},\label{H_h}
\end{equation}
 where
\begin{eqnarray}
\quad H_{\nu}^{(h)} & = & H_{0}^{(\nu_{h})}+H_{int}^{(\nu_{h})}, \\
H_{0}^{(\nu_{h})} & = & \frac{v_{\nu}}{2\pi}\int dy\Big[K_{\nu}(\partial_{y}\theta_{\nu_{h}})^{2}+\frac{1}{K_{\nu}}(\partial_{y}\phi_{\nu_{h}})^{2}\Big]\label{HhBos}   \nonumber \\
H_{int}^{(s_{h})} & = & J^z_\perp \int \frac{\mathtt{d}y}{2\pi \alpha} \cos(4\phi_{s_{h}})\nonumber \\
H_{int}^{(a_{h})} & = & J^{xy}_\perp  \int \frac{\mathtt{d}y}{2\pi \alpha} \cos(2\theta_{a_{\pm}})+J^z_\perp \Lambda \int dy\,\cos(2\phi_{a_{h}}).\nonumber
\end{eqnarray}
The parameters in these expressions are given by
\begin{eqnarray}
 v_{s}\approx v\left(1+\frac{KJ_{\perp}^{z}\alpha}{2\pi v}\right), &\quad & v_{a}\approx v\left(1-\frac{KJ_{\perp}^{z}\alpha}{2\pi v}\right),\label{vKdef}\\
 K_{s}\approx\frac{K}{2}\left(1-\frac{KJ_{\perp}^{z}\alpha}{2\pi v}\right), &\quad& K_{a}\approx2K\left(1+\frac{KJ_{\perp}^{z}\alpha}{2\pi v}\right),\nonumber
\end{eqnarray}
where the original Luttinger liquid parameter $K$ is as given in Eq.~\eqref{Kdef_Delta}.

 This is the standard bosonized form of a spin-$1/2$ ladder \cite{Giamarchi,Shelton,Gogolin}. However,
in the present case due to the spin-charge duality encapsulated by
the helical nature of the modes, $H_{s}^{(h)}$,$H_{a}^{(h)}$ describe
the not just the dynamics of the spin degree of freedom, but also the charge degree of freedom.
To make this concrete, $\partial_{y}\phi_{\nu_{h}}$
denotes spin-density fluctuations as defined in Eq.~\ref{eq:JWtransformation}.  However, it simultaneously encodes the total
(symmetric) and relative (antisymmetric) electric current operators
through channels $1_{h},2_{h}$:
\begin{align}
 & J_{h}^{s}\equiv J_{1_{h}}+J_{2_{h}}=\frac{-2ev h}{\pi K}\partial_{y}\phi_{s_{h}}\,,\label{JaJs}\\
 & J_{h}^{a}\equiv J_{1_{h}}-J_{2_{h}}=\frac{-ev h}{\pi K}\partial_{y}\phi_{a_{h}}\,.\label{JaJs2}
\end{align}
Similarly, the corresponding charge density operators may be written in terms of the dual fields $\partial_{y}\theta_{\nu_{h}}$.  This correspondence will be most useful when we come to calculate conductance properties in Section \ref{sec:Transport-coefficients}.  First however, we must determine the phase diagram of the XXZ spin-ladder model.

\section{\label{sec:Phase-diagram}Phase diagram}

\subsection{Phase boundaries}

In terms of the symmetric and antisymmetric modes, the model is decoupled and so we can treat each of these modes individually.  We begin with the symmetric mode.  This is described by a sine-Gordon model $H_{s}^{(h)}$; the cosine term is relevant (in the renormalisation group sense) when $K_s<1/2$.  Expanding Eqs.~\eqref{vKdef} and \eqref{Kdef_Delta} for small $J^z/J_n^{xy}$ we find
\begin{equation}
K_s \approx \frac{1}{2}\left[ 1 - \frac{J^z_n + J_\perp^z}{2\pi J_n^{xy}}\right].
\end{equation}
This means that the cosine term is relevant if $J_n^z+J_\perp^z>0$ and irrelevant otherwise.  Thus in order to be \textit{irrelevant}, at least one of the magnetic couplings (inter- or intra-chain) must be of a \textit{ferromagnetic} character.  In this case, the symmetric mode remains a gapless Luttinger liquid, and hence the system has low-energy current carrying modes.  We will therefore be most interested in this case when we come to study transport.

In the case that the cosine term is \textit{relevant}, excitations of the symmetric mode acquire a spectral gap, and the ground state is characterised by the locking of the field $\phi_{s_h}$ in one of the minima of the cosine potential.  This means that one of the order parameters $\langle \cos(2\phi_{s_h}) \rangle $ or $\langle \sin(2\phi_{s_h}) \rangle $ gains a non-zero expectation value; the former if $J_\perp^z>0$ and the latter if $J_\perp^z<0$ \cite{Giamarchi,Gogolin}.

We now turn to the antisymmetric mode.  This is described by a Hamiltonian containing two cosine terms:
\begin{eqnarray}
 & H_{a}^{(h)}=\frac{v_{a}}{2\pi}\int dx\Big[K_{a}(\nabla\theta_{a_{h}})^{2}+\frac{1}{K_{a}}(\nabla\phi_{a_{h}})^{2}\Big]\label{H_a}\\
 & +g_{xy}\int \frac{v_a \mathtt{d}y}{2\pi \alpha^2}\cos(2\theta_{a_{h}})+g_{z} \int \frac{v_a \mathtt{d}y}{2\pi \alpha^2} \cos(2\phi_{a_{h}}),\nonumber
\end{eqnarray}
where for convenience we have introduced the dimensionless coupling constants $g_{xy}=J^{xy}_\perp/J^z_n$ and $g_z=J^{z}_\perp/J^z_n$.  In this Hamiltonian, the first cosine term is relevant for $g_{xy}$ has a scaling dimension of $1/K_a$, and is therefore relevant if $K_a>1/2$, while the latter cosine has a scaling dimension of $K_a$ and is therefore relevant if $K_a<2$.  This means firstly that for all values of $K_a$, at least one of the cosine terms is relevant, and therefore one does not expect a gapless Luttinger-liquid like state in the antisymmetric mode; and secondly that for $1/2<K_a<2$ both cosine terms are relevant and compete with each other.

Due to the non-local commutation between the fields $\phi$ and $\theta$, it is not possible for both cosine terms to flow to strong coupling and for both fields to be locked in the ground state.  If only the cosine of the $\theta$ term was present, then a spectral gap $\Delta_a$ would appear
\begin{equation}
\Delta_a \sim \frac{v_a}{\alpha} |g_{xy}|^{\frac{1}{2-1/K_{a}}},
\end{equation}
and one of the expectation values $\langle \cos(\theta_{a_h}) \rangle $ or $\langle \sin(\theta_{a_h}) \rangle $ would be non-zero depending on the sign of $g_{xy}$.  On the other hand, if this term was absent and only the cosine of the $\phi$ term was present, then the spectral gap would be given by
\begin{equation}
\Delta_a \sim \frac{v_a}{\alpha} |g_{z}|^{\frac{1}{2-K_{a}}},
\end{equation}
and the non-zero expectation values would be one of $\langle \cos(\phi_{a_h}) \rangle $ or $\langle \sin(\phi_{a_h}) \rangle $ depending on the sign of $g_z$.

When both terms are present, by very general arguments one expects the stronger term to 'win', and therefore the location of the phase boundary between these two possibilities may be estimated as the point where the mass that would be generated by each of the terms individually becomes equal (see e.g. Ref.~\onlinecite{Sam_even_older})
\begin{equation}
|g_{xy}|^{\frac{1}{2-1/K_{a}}}\sim|g_{z}|^{\frac{1}{2-K_{a}}}\;.\label{QCPvsK}
\end{equation}
It is possible to confirm this picture by a more formal treatment \cite{Shelton,Gogolin,SDSG,Atzmon} based on refermionization around the self-dual point $K_a=1$.  From this method, one furthermore deduces that the phase transition between ordered $\phi$ and ordered $\theta$ is in the Ising ($Z_2$) universality class.  We refer to Ref.~\cite{Mazo-1} for an outline of the derivation of this as applied to this particular model.

\subsection{Local operators and nature of phases}

Having determined the location and nature of the phase boundaries in the model, it still remains to determine the nature of each of the phases.  As the model may be written in either spin or charge language, the phases can also be characterized from either of these viewpoints.  We begin with the spin-ladder picture.  In order to characterize the phases, we must find local (spin) operators that either gain an expectation value (long range order), or in some cases have the slowest decaying power-law correlation functions (quasi long range order).  In order to do this, we define four operators on the rungs of the ladders
\begin{equation}
{\cal O}^{x,z}_\pm = (-1)^y \left[S_1^{x,z} \pm S_2^{x,z} \right].
\end{equation}
In words, these correspond to rung ferromagnetism ($+$) or antiferromagnetism ($-$) in the $x$ or $z$ spin direction, and we are interested in the staggered (antiferromagnetic) component in the chain direction.  By applying the transformations \eqref{eq:JWtransformation} and \eqref{eq:rotation}, we find the bosonized forms of these operators to be
\begin{eqnarray}
{\cal O}_+^z  &\sim& \cos 2\phi_s \cos \phi_a, \nonumber \\
{\cal O}_-^z  &\sim& \sin 2\phi_s \sin \phi_a, \nonumber \\
{\cal O}_+^x  &\sim& \cos (\theta_s/2) \cos \theta_a, \nonumber \\
{\cal O}_-^x  &\sim& \sin (\theta_s/2) \sin \theta_a.
\label{eq:localops}
\end{eqnarray}

\begin{figure}
\begin{centering}
\includegraphics[width=1\linewidth]{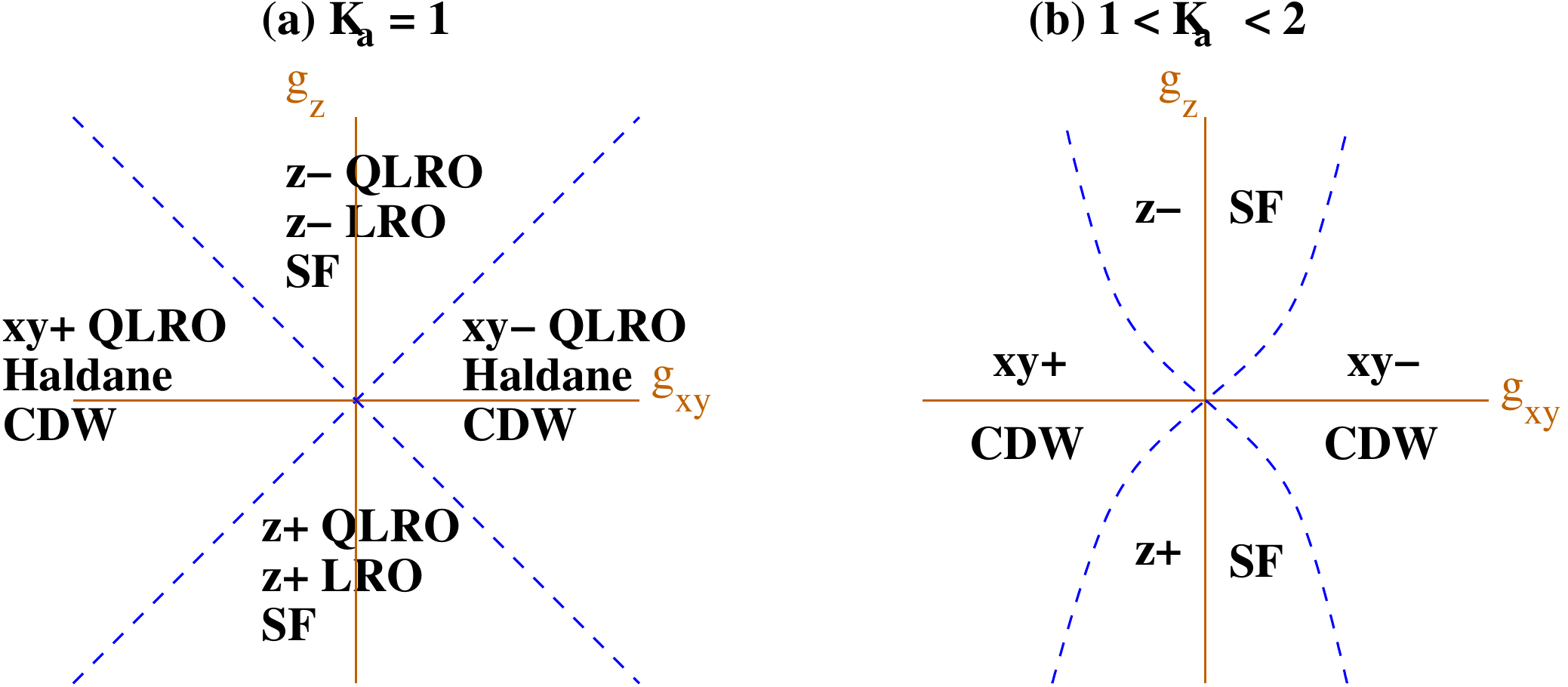}
\caption{Phase diagram of the Spin ladder as a function of the inter-chain couplings $g_z$ and $g_{xy}$.  The QLRO order states refer to the case when $K_s>1/2$ and the symmetric modes are gapless.  The LRO/Haldane labels refer to the case $K_s<1/2$ and the symmetric modes are gapped.  The SF/CDW is the equivalent designation of the state from the charge point of view.  The correlations in the ground states corresponding to each of the labels is described in the main text.  Part (a) shows the phase diagram when other parameters are tuned such that $K_a=1$.  If $1<K_a<2$, the transition lines become curved, as shown in part (b).  If $K_a>2$, then the $z\pm$/CDW phases disappear completely.  We note that as a function of model parameters, these diagrams are only schematic as the Luttinger parameters $K$ also depend on the inter-chain couplings.  The general expression for the location of the phase boundaries is given in the main text in Eq.~\ref{QCPvsK}. }\label{fig:PhaseDiagramS}
\end{centering}
\end{figure}

We can now discuss the phase diagram of the model.  To be concrete, we consider the case when $K_a=1$, so that the phase transitions occur at $|g_{xy}|=|g_z|$ (see Eq.~(\ref{QCPvsK})).  We begin by discussing the case when $K_s>1/2$ so the symmetric modes remain gapless.  In this case, none of the above operators can have a ground state expectation value, as they all involve the gapless symmetric modes.  The phases in such a system are then characterised by the operators with the slowest decaying correlation functions, known as quasi-long-range order (QLRO)\cite{Giamarchi,Sam_short,Sam_long}.  We see therefore that if $|g_z|>|g_{xy}|$  this operator is ${\cal O}_-^z$ if $g_z>0$ (for simplicity we call this state $z-$ QLRO); while if $g_z<0$ then the state is $z+$ QLRO.  This is easy to understand: if the legs are coupled by an easy-axis exchange term, then the relative spin configuration on each leg will reflect the sign of this exchange.  Similarly, in the opposite case if $|g_{xy}|>|g_z|$, then one has $xy-$ or $xy+$ QLRO depending on the sign of $g_{xy}$.  We refer to these states as $xy\pm$ rather than just $x\pm$ as the correlation functions in the $y$ spin projection (which we haven't explicitly written) follow the same power-law decay as those in the $x$.  This is related to the $O(1)$ symmetry of the model, we will come back to this point later.
This phase diagram is plotted in Fig.~\ref{fig:PhaseDiagramS}, along with the equivalent phase diagram for a different value of $K_a$.

We now turn to the case when $K_s<1/2$, i.e. both the symmetric and antisymmetric modes have gaps, and again we start with the case $|g_z|>|g_{xy}|$.  As the same coefficient $g_z$ appears in front of the cosine terms in both modes, it is easy to see that one finds $\langle {\cal O}_-^z \rangle \ne 0$ if $g_z>0$ and $\langle {\cal O}_+^z \rangle \ne 0$ if $g_z<0$.  In other words, the $z\pm$ QLRO states previously described have now acquired true long range order.  These states correspond to spin arrangements of an Ising antiferromagnet along the legs, with the two legs either in phase ($z+$) or out of phase ($z-$) depending on the sign of the interchain coupling.  These states are associated with a spontaneously broken $Z_2$ symmetry, bringing further insight into the Ising nature of the phase transition previously discussed.

The same does not happen for the $x\pm$ states.  It is easy to see why: an expectation value of one of these operators would imply breaking of a continuous $O(1)$ symmetry, which is forbidden in one dimension.   In fact, the operators ${\cal O}_\pm^x$ which exhibited slow power law decay for $K_s>1/2$ will show exponentially decaying correlations at $K_s<1/2$.  This is technically because these operators contain cosines of the $\theta_s$ field, but the strong coupling ground state in the symmetric modes are associated with locking of the $\phi$ fields.  It is possible to show \cite{Shelton} that no local operator can be associated with the combination of the $\phi_s$ and $\theta_a$ fields; the expectation value of these instead refers to a \textit{string order parameter}, associated with the spontaneous breaking of a $Z_2$ topological symmetry.  This is the celebrated Haldane phase.  These phase labels are also shown in the phase diagram in Fig.~\ref{fig:PhaseDiagramS}.

Having now described the phase diagram of the spin-ladder, we can ask what these different phases would correspond to in terms of charge modes, coupled to the spin modes by the helical nature of the ladder.  Going back to Eq.~(\ref{JaJs2}), we see that the gradient of the $\phi$ fields are associated with the current.  By the general principles on one-dimensional physics \cite{Giamarchi}, this means that the $\phi$ fields are associated with the phase of fermionic operators -- meaning that a locking of $\phi_a$ may be associated with locking of the relative \textit{charge phase} between the legs, characteristic of a superfluid (SF).  On the other hand, $\theta$ is associated with the charge degrees of freedom and hence phases with locked charge may be characterised as charge density waves (CDW).  Thus the phases labeled $z\pm$ in the spin language are SF in terms of charge degrees of freedom, while the $xy\pm$ phases are CDWs.  These labels are also shown in the phase diagram in Fig.~\ref{fig:PhaseDiagramS}; it is this coupling between spin phases and charge phases that allow charge transport experiments to probe the different spin phases of the spin ladder, as we will demonstrate in the next section.

\section{\label{sec:Transport-coefficients}Transport coefficients}
\label{Conductance}

To derive the conduction properties characterizing the distinct phases,
we first introduce local coupling terms between the channels $1_{h},2_{h}$
which break translation invariance in the $y$-direction, and are
necessary to induce non-trivial transport coefficients. As a minimal
choice of such terms, we consider defects
at $y=0$ which add a local correction $J_{0}$ to $J_{\perp}^{xy}$
{[}Eq. (\ref{spinladder}){]}, and a spin-flip term allowing backscattering
between the closest channels of opposite helicities {[}$H_{+-}$ in
Eq. (\ref{2ladders}){]}:
\begin{align}
 & \delta H^{(h)}=J_{0}\big[S_{1_{h},0}^{+}S_{2_{h},0}^{-}+h.c.\big]\,,\label{H_local}\\
 & H_{+-}=J\big[S_{1_{-},0}^{+}S_{1_{+},0}^{-}+h.c.\big].\nonumber
\end{align}
 In terms of Bosonic fields, these yield
\begin{eqnarray}
\delta H & = & \sum_{h=\pm}J_{0}\Lambda\cos[\theta_{1_{h}}(0)-\theta_{2_{h}}(0)]\label{H_local_Bos}\\
 & + & \sum_{n,n^{\prime}=1,2}J_{n,n^{\prime}}\Lambda\cos[\theta_{n_{+}}(0)-\theta_{n_{-}^{\prime}}(0)]\nonumber
\end{eqnarray}
 where $J_{n,n^{\prime}}$ with $n,n^{\prime}=2$ are generated to
second order in the perturbations Eq. (\ref{H_local}).  While this minimal model involves a single impurity (defect) in the system, the arguments can be easily extended to a finite density of impurities \cite{Giamarchi}, and would yield qualitatively similar results.

\begin{figure}
\begin{centering}
\includegraphics[width=1\linewidth]{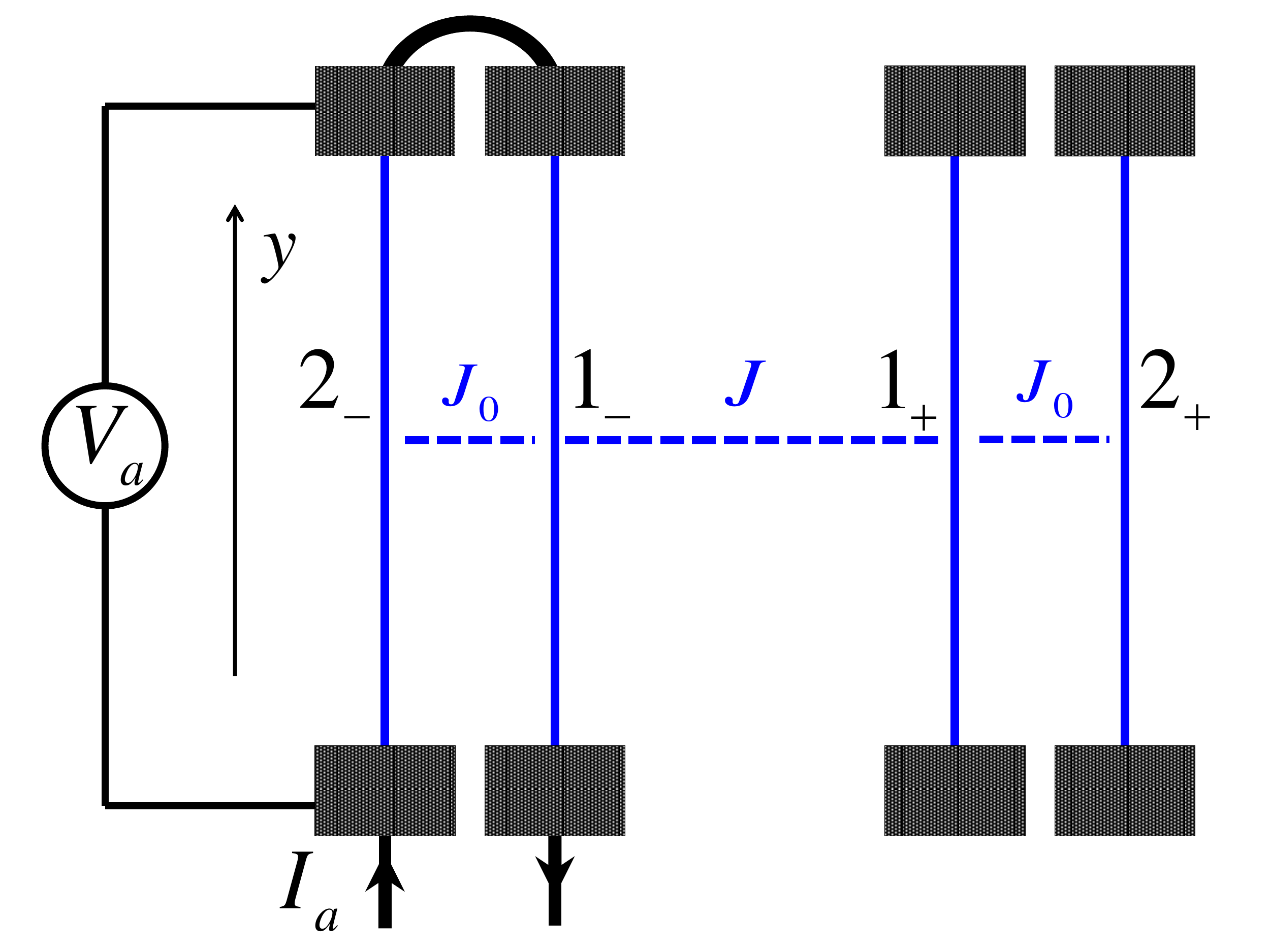} \caption{Schematic transport measurement geometry, illustrated for antisymmetric
conductance. Blue solid lines represent DW's and $J$, $J_{0}$ are
defined in Eq. (\ref{H_local}).\label{fig:G_a}}
\par\end{centering}
\centering{}
\end{figure}

We now consider a multi-terminal contact to an external circuitry
where current can be driven along the $y$-direction of the BLG sample
(Fig.~\ref{fig:G_a}). We particularly focus below on two observables:
the total two-terminal conductance $G$, and the ``antisymmetric
conductance\textquotedbl{} $G_{a}=I_{a}/V_{a}$ where $I_{a}$ is
a counter-propagating current in the channels $1_{h},2_{h}$, short-circuited
at one edge (see Fig. \ref{fig:G_a}).  We will show that either of these conductance measurements is extremely sensitive to the ground state of the spin-ladder model, and in particular the dependence of conductance on temperature is qualitatively different depending on whether the system is in the SF ($z-$QLRO in the spin-ladder language) or the CDW ($xy-$QLRO) state.  In this section, we limit ourselves to these two cases, which occur when the symmetric mode remains gapless ($K_s>1/2$).  A comparison of the conduction properties of the other states, which are fully gapped and therefore non-metallic, will be presented elsewhere.

\subsection{Total Conductance}

From Kubo's formula, $G$ is given by the retarded correlation function
of the fully symmetric current
\begin{equation}
J_{s}=\sum_{h=\pm}J_{h}^{s},\label{J_s}
\end{equation}
 where $J_{h}^{s}$ is given in Eq. (\ref{JaJs}). We consider the
behavior of $G(T)$ at a finite temperature $T$ under the assumption
of weak backscattering between the channels of opposite helicity.
The main contribution to the scattering of the current $J_{s}$ arises
from the second term in Eq. (\ref{H_local_Bos}), which couples the
$a$ and $s$ modes via the operators
\begin{equation}
{\cal O}_{\pm}=\cos\left[\frac{\theta_{s_{+}}(0)-\theta_{s_{-}}(0)}{2}\right]\cos\left[\theta_{a_{+}}(0)\pm\theta_{a_{-}}(0)\right]\;.\label{H+_as}
\end{equation}
 To leading order in $\delta H$ ($\propto J^{2}$, see Eq. (\ref{H_local})
and Fig. \ref{fig:G_a}), the conductance $G$ (in units of $e^{2}/2\pi\hbar$)
is then given by \cite{Chap7}
\begin{eqnarray}
G & = & 4-\delta G\nonumber \\
\delta G & \sim & \int_{0}^{\infty}dt\, t\langle[F_{\pm}(t),F_{\pm}(0)]\rangle_{0},\label{G_Kubo}
\end{eqnarray}
 where
\begin{eqnarray}
 & \quad F_{\pm}\equiv i[J_{s},{\cal O}_{\pm}]\;.
\end{eqnarray}
 Here, the expectation value $\langle...\rangle_{0}$ is evaluated
with respect to $H_{0}=\sum_{h=\pm}H_{h}$, where $H_{h}$ is given
in Eq. (\ref{H_h}). The resulting $\delta G(T)$ depends on the behavior
of the correlation function in Eq. (\ref{G_Kubo}), which is distinct
in the two phases. We therefore discuss the CDW and SF phases separately.

%\subsubsection*{CDW phase}

\begin{figure}[h]

\begin{centering}
\includegraphics[width=1\linewidth]{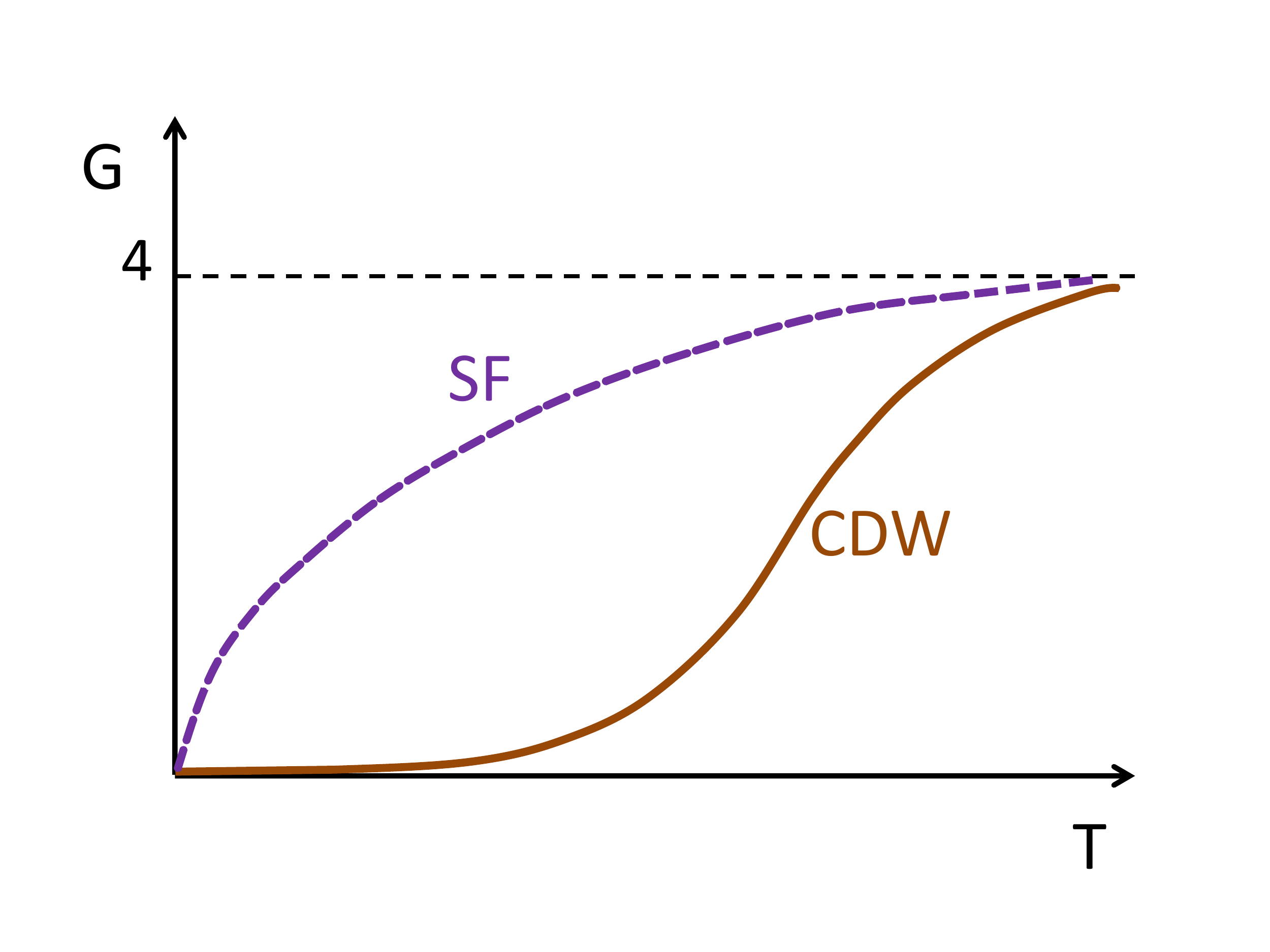} \includegraphics[width=1\linewidth]{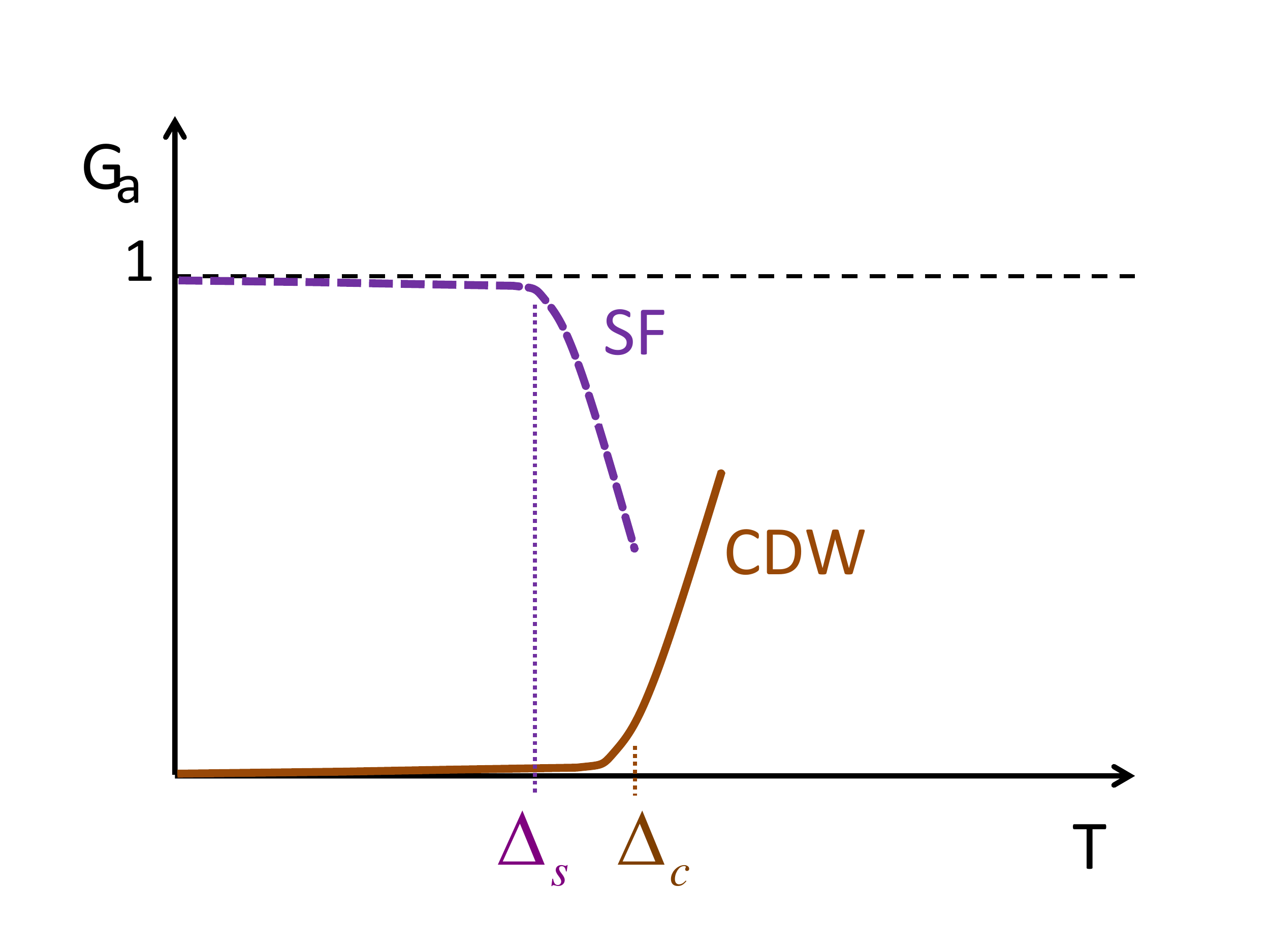}
\caption{(color online). Qualitative behavior of the transport coefficients
as functions of $T$ in the two phases: solid brown curves correspond
to the CDW phase, and dashed purple curves to the SF phase. Top panel:
the total conductance $G$. The high-$T$ regime is dominated by $G=4-\delta G$,
with $\delta G$ given by Eqs. (\ref{G_highT}) and Eq. (\ref{G_SF_highT})
in the CDW and SF phases, respectively. The low-$T$ regime is dominated
by Eqs. (\ref{G_I_lowT}) and (\ref{G_SF_lowT}) in the CDW and SF
phases, respectively. Bottom panel: the antisymmetric conductance
$G_{a}$, given by Eqs. (\ref{Ga2deltaGa}), (\ref{G_a_S}) in the
SF phase and Eq. (\ref{G_a_I}) in the CDW phase.\label{fig:conductances}}

\par\end{centering}

\centering{}
\end{figure}

In the CDW phase, since $\theta_{a_{h}}$ are ordered, the second
cosine in Eq. (\ref{H+_as}) can be replaced by its finite expectation
value $\sim1$ and one obtains
\begin{eqnarray}
 & {\cal O}_{\pm}\sim\cos\theta\,,\\
 & \text{with}\quad\theta\equiv\frac{1}{2}\big(\theta_{s_{+}}(0)-\theta_{s_{-}}(0)\big)\;.\nonumber
\end{eqnarray}
 The total current $J_{s}$ can be expressed in terms of the the canonically
conjugate field $\phi\equiv\phi_{s+}-\phi_{s-}$: from Eqs. (\ref{JaJs}),(\ref{J_s}),
\[
J_{s}=-\frac{ev}{\pi K}\partial_{y}\phi\,.
\]
 Employing Eq. (\ref{G_Kubo}), and recalling that the s-mode is a
Luttinger liquid, we get a power-law $T$-dependence
\begin{eqnarray}
 & \delta G(T)\sim T^{\frac{1}{4K_{s}}-2}\;.\label{G_highT}
\end{eqnarray}
 For accessible values of $K_{s}$, this typically \textit{diverges}
at low $T$, implying a breakdown of the weak backscattering approximation
and Eq. (\ref{G_highT}) is no longer valid. The system therefore
exhibits an insulating behavior, $G(T\rightarrow0)=0$. The finite
low-$T$ dependence of the total conductance can be evaluated perturbatively
as a correlation function of the dual tunneling operator \cite{KaneFisher,Saleur,Sam_short,Sam_long}
\begin{eqnarray}
 & {\cal O}^{(d)}\sim\cos(4\phi).
\end{eqnarray}
 This yields (for $T\ll T_{c}=v\Lambda$)
\begin{equation}
G(T)\sim T^{16K_{s}-2}\;.\label{G_I_lowT}
\end{equation}
 The overall $T$-dependence of the total conductance $G(T)$ in the
CDW phase, interpolating between the high and low $T$ regimes (Eqs.
(\ref{G_highT}) and (\ref{G_I_lowT}), respectively), is sketched
as a solid brown curve in the top panel of Fig. \ref{fig:conductances}.

%\subsubsection*{SF phase}

In the SF phase, $\theta_{a_{h}}$ are disordered and the correlations
of $e^{\pm i\theta_{a_{h}}}$ yield an exponential decay of $\delta G$
evaluated from Eq. (\ref{G_Kubo}) for $T\ll\Delta_{s}$ (with $\Delta_{s}$
the SF gap). The leading backscattering is therefore governed by \textit{second}
order terms generated by $\delta H$, which decouple the $a$-mode
\cite{OG,Atzmon,Sam_short,Sam_long}, of the form
\begin{eqnarray*}
 & {\cal O}_{\pm}\sim\cos(2\theta)\;.
\end{eqnarray*}
 One therefore obtains
\begin{equation}
\delta G\sim T^{\frac{1}{K_{s}}-2}\;,\label{G_SF_highT}
\end{equation}
 which under our assumption $K_{s}>1/2$ still implies an \textit{insulating}
behavior. The same procedure leading to Eq. (\ref{G_I_lowT}) can
be implemented, evaluating $G(T)$ perturbatively in the dual tunneling
operator which now takes the form
\begin{equation}
{\cal O}^{(d)}\sim\cos(2\phi)\;.
\end{equation}
 This yields for finite low $T$
\begin{equation}
G(T)\sim T^{4K_{s}-2}\label{G_SF_lowT}
\end{equation}
 (which approaches a non-universal constant for $K_{s}\sim1/2$).
The overall $T$-dependence of the total conductance $G(T)$ in the
SF phase, interpolating between the high and low $T$ regimes (Eqs.
(\ref{G_SF_highT}) and (\ref{G_SF_lowT}), respectively), is sketched
as a dashed purple curve in the top panel of Fig. \ref{fig:conductances}.
Comparing the low-$T$ behavior of $G(T)$ in the two phases, we conclude
that a transition from SF to CDW is manifested as a jump in the power-law
$G\sim T^{\kappa}$, from $\kappa=4K_{s}-2$ to $\kappa=16K_{s}-2$.

\subsection{Antisymmetric Conductance}

A more dramatic signature of the SF/CDW transition is expected in
the $T$-dependence of $G_{a}$, which probes the response to a pure
antisymmetric current $I_{a}$. Backscattering in this channel is
solely due to the first term in Eq. (\ref{H_local_Bos}), which can
be cast as
\begin{equation}
{\cal O}_{a}=J_{0}\Lambda\sum_{h=\pm}\cos\left[2\theta_{a_{h}}(0)\right]\;.\label{O_a}
\end{equation}
 $G_{a}$ (for each ladder $h=\pm$) is dictated by the correlation
of the relative current operators $J_{h}^{a}$ {[}Eq. (\ref{JaJs2}){]}.

%\subsubsection*{SF phase}

We first consider the SF phase, where the deviation $\delta G_{a}$
from perfect conductance
\begin{equation}
G_{a}=1-\delta G_{a}\label{Ga2deltaGa}
\end{equation}
 can be evaluated perturbatively from Eq. (\ref{G_Kubo}), with ${\cal O}_{\pm}$
replaced by ${\cal O}_{a}$, associated with the disordered operators
in this phase. For $T\ll\Delta_{s}$, this yields
\begin{equation}
\delta G_{a}\sim\exp\left(-\frac{\Delta_{s}}{T}\right)\;.\label{G_a_S}
\end{equation}
 This implies an exponentially small voltage drop $V_{a}\sim\delta G_{a}$
in the setup depicted in Fig. \ref{fig:G_a}.

In the CDW phase, the leading contribution to the antisymmetric conductance
$G_{a}$ (which tends to vanish in the $T\rightarrow0$ limit) can
be obtain following a similar calculation, with ${\cal O}_{a}$ replaced
by its dual
\begin{equation}
{\cal O}_{a}^{(d)}\sim\sum_{h}\cos\left[2\phi_{a_{h}}(0)\right]\;.\label{O_ad}
\end{equation}
 This yields an exponentially small expression for the antisymmetric
conductance:
\begin{equation}
G_{a}\sim\exp\left(-\frac{\Delta_{c}}{T}\right)\label{G_a_I}
\end{equation}
 with $\Delta_{c}$ a charge gap characterizing this phase.

The overall behavior of $G_{a}$ in the low-$T$ regime is sketched
in the bottom panel of Fig. \ref{fig:conductances}. This figure summarizes
our main prediction: $G_{a}$ would exhibit a true ``superconductor-insulator\textquotedbl{}
transition, indicated by a jump of $G(T\rightarrow0)$ from 1 to 0
upon tuning of, e.g. $J_\perp^{z}/J_\perp^{xy}$, through the phase boundaries of Fig.
\ref{fig:PhaseDiagramS}. Alternatively, for fixed $J_\perp^{z}/J_\perp^{xy}$ the transition can be accessed 
by varying the Luttinger parameter $K_a$.
Since $K_a$ monotonically increases with
the physical parameter $\ell V/e \propto V/\sqrt{B_{z}}$ (see Eq.~\ref{eq:anisotropy}), the transition
can in principle be observable by continuous tuning of either the
electric or magnetic fields.

\section{Conclusions}

\label{Conclusions}

To summarize, we propose a designed realization of quantum spin ladders in an electronic setup where a suspended BLG is non-uniformly gated
by split double-gates, and subject to a tilted magnetic field ${\bf B}$.  When the split gates have opposite polarities (as depicted in Fig. \ref{fig:dgated_BLG}), a kink is created in the perpendicular electric field at the middle of the sample. We have shown that in the $\nu=0$ QH state
of the BLG, this leads to the formation of two pairs of coupled DW modes at opposite sides of the kink. The quantum dynamics of each pair can be modeled as an anisotropic spin-$1/2$ two-leg ladder, where due to helicity of the underlying electronic states the spin degree of freedom is linked to the charge.

Thus, unlike spin-ladder compounds found in nature, this realization allows one to isolate a single ladder with truly 1D dynamics. Moreover, their exchange parameters are highly tunable by changing the strength and tilt-angle of ${\bf B}$, the gate-voltage $V$ and the separation $w$ of the split gates. Most prominently, due to the spin-charge coupling, distinct phases of the spin sector possess very different charge transport properties.
In particular, we identify an Ising-like transition between two phases.  In the spin language, these phases are characterised by leading power-law decays in either the $z$ spin-direction or in the $xy$ plane.  In terms of charge however, these same phases may be labelled as
an insulating CDW and a SF. Their distinct character is then most dramatically manifested by the temperature dependence of
an antisymmetric conductance $G_{a}$ (see bottom panel of Fig. \ref{fig:conductances}).

The above predicted superfluid-insulator transition should be observable in experimental conditions where the
Luttinger parameter $K_{a}$ appearing in our model is of order 1. While the detailed relation of this parameter to physical parameters of the apparatus is complicated, a rough estimate associates it with the ratio of Coulomb energy $e^2/\ell$ and the kinetic energy set by the gate voltage, $eV$. The desired regime of parameters therefore corresponds to $e/\ell V\sim 1$, implying that, e.g., for $B_{z}\sim0.1$T one should apply a perpendicular electric field $E_{\perp}\sim100\frac{{\rm mV}}{{\rm nm}}$. This is within the range of parameters accessible in several leading labs (see, for example, \cite{Yacoby2011}). An additional requirement is that the Zeeman energy (determined by $|{\bf B}|$) is sufficiently strong to generate a sufficient splitting $d$ between the two pairs of ladders [see Eq. (\ref{eq:d})]. This can be achieved in strong ${\bf B}$ where $|{\bf B}|>10$T but the tilt-angle is sufficiently large such that $B_z\ll |{\bf B}|$; see, e.g., the experimental setup of Ref. \cite{Young2013}.

%As a final remark, note that the paired DW's discussed in our case
%are apparently analogous to a ``helical ladder\textquotedbl{} formed
%by coupling two parallel edge states of TI, with a crucial distinction:
%in the latter, a coupling in the form of the first term in $H_{\perp}^{(h)}$
%{[}Eq. (\ref{spinladder}){]} is forbidden. The second term of $H_{\perp}^{(h)}$,
%analogous to a Josephson coupling resulting from electron-pair tunneling
%between the HLL's, likewise competes with a term generating a CDW
%order. However, the latter has a different scaling dimension. Hence,
%the resulting SF-CDW transition is of a different nature. This will
%be discussed in more detail elsewhere \cite{future}.

We thank E. Berg, T. Pereg-Barnea, A. Stern and A. Young for useful discussions.
E. S. is grateful to the hospitality of the Aspen Center for Physics
(NSF Grant No. 1066293) and to the Simons Foundation. This work was
supported by the US-Israel Binational Science Foundation (BSF) grant
2012120, the Israel Science Foundation (ISF) grant 599/10, and by
NSF Grant No. DMR-1005035.

\expandafter\ifx\csname natexlab\endcsname\relax\global\long\def\natexlab#1{#1}
\fi \expandafter\ifx\csname bibnamefont\endcsname\relax \global\long\def\bibnamefont#1{#1}
\fi \expandafter\ifx\csname bibfnamefont\endcsname\relax \global\long\def\bibfnamefont#1{#1}
\fi \expandafter\ifx\csname citenamefont\endcsname\relax \global\long\def\citenamefont#1{#1}
\fi \expandafter\ifx\csname url\endcsname\relax \global\long\def\url#1{\texttt{#1}}
\fi \expandafter\ifx\csname urlprefix\endcsname\relax\global\long\def\urlprefix{URL }
\fi \providecommand{\bibinfo}[2]{#2} \providecommand{\eprint}[2][]{\url{#2}}

\end{document}